\headline={\ifnum\pageno>1 \hss \number\pageno\ \hss \else\hfill \fi}
\pageno=1
\nopagenumbers

\centerline{ \bf THE HIGHER COHOMOLOGIES OF $E_8$ LIE ALGEBRA}
\vskip 15mm
\centerline{\bf H. R. Karadayi and M. Gungormez}
\centerline{Dept.Physics, Fac. Science, Tech.Univ.Istanbul }
\centerline{ 80626, Maslak, Istanbul, Turkey }

\vskip 10mm
\centerline{\bf{Abstract}}
\vskip 10mm
It is well known that anomaly cancellations for $D_{16}$ Lie algebra are at
the root of the first string revolution. For $E_8$ Lie algebra, cancellation
of anomalies is the principal fact leading to the existence of heterotic
string. They are in fact nothing but the 6th order cohomologies of
corresponding Lie algebras. Beyond 6th order, the calculations seem to
require special care and it could be that their study will be worthwhile
in the light of developments of the second string revolution.

As we have shown in a recent article, for $A_N$ Lie algebras, there is a
method which are based on the calculations of Casimir eigenvalues. This is
extended to $E_8$ Lie algebra in the present article. In the generality of
any irreducible representation of $E_8$ Lie algebra, we consider 8th and
12th order cohomologies while emphasizing the diversities between the two.
It is seen that one can respectively define 2 and 8 basic invariant polinomials
in terms of which 8th and 12th order Casimir eigenvalues are always expressed
as linear superpositions. All these can be easily investigated because
each one of these invariant polinomials gives us a linear equation
to calculate $E_8$ weight multiplicities. Our results beyond order 12 are
not included here because they get more complicated though share the same
characteristic properties with 12th order calculations.
\footnote{}{e-mail: karadayi@sariyer.cc.itu.edu.tr}

\vskip 15mm
\vskip 15mm
\vskip 15mm
\vskip 15mm
\vskip 15mm
\vskip 15mm

\hfill\eject

\vskip 3mm
\noindent {\bf{I.\ INTRODUCTION}}
\vskip 3mm
It is a clear fact that anomaly cancellations play a unique role in the
construction of the way of thinking and constructing models in high energy
physics since the last two decades. The ones for $D_{16}$ Lie algebra {\bf [1]}
are principal for the first string revolution to begin. As it is also
noted {\bf [2]}, the contruction of heterotic string {\bf [3]} is {\bf shortly
thereafter}. It is known {\bf [4]} that the existence of a 10-dimensional
string with a $E_8 \times E_8$ gauge symmetry relies heavily on $E_8$ anomaly
cancellations.

On the other hand, these anomaly cancellations are in fact due to cohomology
relations of corresponding Lie algebras. The cohomology for Lie algebras states
non-linear relationships between elements of the center of their universal
enveloping algebras {\bf [5]}. The non-linearity comes from the fact that these
relationships are between the elements of different orders and the non-linearly
independent ones are determined by the Betti numbers {\bf [6]}. A problem
here is to determine the number of linearly independent elements of the same
order. In two subsequent works {\bf [7]}, we studied this problem for $A_N$ Lie
algebras and give a method which is based on explicit construction of Casimir
eigenvalues. This will be extended here to $E_8$ Lie algebra.

$E_8$ is the biggest one of finite dimensional Lie algebras and besides its
own mathematical interest it plays a striking role in high energy physics.
It provides a natural laboratory to study the structure of $E_{10}$ hyperbolic
Lie algebra {\bf [8]} which is seen to play a key role in understanding the
structure of infinite dimensional Lie algebras beyond affine Kac-Moody Lie
algebras. There are so much works to show its significance in string theories
and in the duality properties of supersymmetric gauge theories. This hence
could give us some insight to calculate higher order cohomologies of $E_8$
Lie algebra. It will be seen in the following that this task is to be
simplified to great extent when one uses a method based on explicit
calculations of Casimir eigenvalues.

It is known that, beside degree 2, $E_8$ Betti numbers give us non-linearly
independent Casimir elements for the degrees  8,12,14,18,20,24,30. We must
therefore calculate the Casimir eigenvalues for all these degrees. In the
present state of work, to give only the results for 8th and 12th orders would
be more instructive. This will be possible in terms of one of the maximal
subalgebras of $E_8$, namely $A_8$. Although our method {\bf [7]} for $A_N$
Lie algebras is previously presented, the calculations still need some special
care for 8th and 12th orders. These are investigated in sections II and III.
To this end, we especially emphasize our second permutational lemma to
express the weights of an $E_8$ Weyl orbit and $A_8$ duality rules without
which the calculations will be useless. In section IV, we show that the
calculations find an end in the form of decompositions in terms of some
properly chosen $A_8$ basis functions . The remarkable fact here is that
the coefficients in these decompositions are constants and this shows us
that the dependence on irreducible representations of $E_8$ Lie algebra
are contained in these $A_8$ basis functions solely. For 12th order, the
results of our calculations are given in three appendices because they are
comparatively voluminous than 8th order calculations.

\vskip 3mm
\noindent {\bf{II.\ WEIGHT CLASSIFICATION OF $E_8$ WEYL ORBITS}}
\vskip 3mm
We refer the excellent book of Humphreys {\bf [9]} for technical aspects of
this section though a brief account of our framework will also be given here.
It is known that the weights of an irreducible representation $R(\Lambda^+)$
can be decomposed in the form of
$$ R(\Lambda^+) = \Pi(\Lambda^+) \ \ + \sum m(\lambda^+ < \Lambda^+) \ \
\Pi(\lambda^+) \eqno(II.1) $$
where $\Lambda^+$ is the principal dominant weight of the representation,
$\lambda^+$'s are their sub-dominant weights and $m(\lambda^+ < \Lambda^+)$'s
are multiplicities of weights $\lambda^+$ within the representation
$ R(\Lambda^+) $. Once a convenient definition of eigenvalues is assigned
to $ \Pi(\lambda^+) $, it is clear that this also means for the whole
$R(\Lambda^+)$ via (II.1).

In the conventional formulation, it is natural to define Casimir eigenvalues
for irreducible representations which are known to have matrix representations.
In ref(7), we have shown that the eigenvalue concept can be conveniently
extended to Weyl orbits of $A_N$ Lie algebras. The convenience comes from a
permutational lemma governing $A_N$ Weyl orbits. This however could not be
so clear for Lie algebras other than $A_N$. We therefore give in the following
a second permutational lemma. To this end, it is useful to decompose $E_8$
Weyl orbits in the form of
$$ \Pi(\lambda^+) \equiv  \sum_{\sigma^+ \in \Sigma(\lambda^+)} \Pi(\sigma^+)
\eqno(II.2) . $$
where
\vskip 3mm
\noindent {\bf $\Sigma(\lambda^+)$ is the set of $A_8$ dominant weights
participating within the same $E_8$ Weyl orbit $\Pi(\lambda^+)$ . }
\vskip 3mm
\noindent If one is able to determine the set $\Sigma(\lambda^+)$ completely,
the weights of each particular $A_8$ Weyl orbit $\Pi(\sigma^+)$ and hence the
whole $\Pi(\lambda^+)$ are known. We thus extend the eigenvalue concept to
$E_8$ Weyl orbits just as in the case of $A_N$ Lie algebras.

It is known, on the other hand, that elements of $\Sigma(\lambda^+)$ have the
same square length with the $E_8$ dominant weight $\lambda^+$. It is
unfortunate that this remains unsufficient to obtain the whole structure of
the set $\Sigma(\lambda^+)$. This exposes more severe problems especially for
Lie algebras having Dynkin diagrams with higher degree automorphisms,
for instance affine Kac-Moody algebras. To solve this non-trivial part
of this problem, we introduce 9 fundamental weights $\mu_I$ of $A_8$, via
scalar products
$$ \kappa(\mu_I,\mu_J) \equiv \delta_{IJ} - {1 \over 9} \ \ , \ \
I,J = 1,2, .. 9 \eqno(II.3)  $$
The existence of $\kappa(.,.)$ is known to be guaranteed by $A_8$ Cartan matrix.
The fundamental dominant weights of $A_8$ are now expressed by
$$  \sigma_i \equiv \sum_{j=1}^{i} \mu_j \ \ , \ \ i=1,2, .. 8. \eqno(II.4)  $$
To prevent misconception, we list the main quantities which take place in the
following discussions:
\vskip 3mm
\leftline{ \ \ \ \ \ \ \ \ \ \ \ \ \ \ \ \ \ \ \ \ $ \lambda^+ \ , \ \Lambda^+
\longrightarrow $ dominant weights of $ E_8 $ }

\leftline { \ \ \ \ \ \ \ \ \ \ \ \ \ \ \ \ \ \ \ \ \ \ \ \ \ \ \ \
$\lambda_i \longrightarrow $ fundamental dominant weights of $ E_8 $ \ , \ i=1,2, .. 8 }

\leftline { \ \ \ \ \ \ \ \ \ \ \ \ \ \ \ \ \ \ \ \ \ \ \ \ \ \ \ $ \sigma^+ \longrightarrow $ dominant weights of $ A_8 $ }

\leftline { \ \ \ \ \ \ \ \ \ \ \ \ \ \ \ \ \ \ \ \ \ \ \ \ \ \ \ \
$ \sigma_i \longrightarrow $ fundamental dominant weights of $A_8$ \ , \ i=1,2, .. 8 }

\leftline {\ \ \ \ \ \ \ \ \ \ \ \ \ \ \ \ \ \ \ \ \ \ \ \ \ \ \ \ \
$ \mu_I \longrightarrow $ fundamental weights of $ A_8 $ \ , \ I=1,2, ... 9 }
\vskip 3mm
\noindent The correspondence $E_8$ $\leftrightarrow$  $A_8$ is now provided by
$$ \eqalign{
\lambda_1 &= \sigma_1 + \sigma_8 \cr
\lambda_2 &= \sigma_2 + 2 \ \sigma_8  \cr
\lambda_3 &= \sigma_3 + 3 \ \sigma_8  \cr
\lambda_4 &= \sigma_4 + 4 \ \sigma_8  \cr
\lambda_5 &= \sigma_5 + 5 \ \sigma_8  \cr
\lambda_6 &= \sigma_6 + 3 \ \sigma_8  \cr
\lambda_7 &= \sigma_7 + \sigma_8  \cr
\lambda_8 &= 3 \ \sigma_8  }   \eqno(II.5)    $$
with
$$ \lambda^+ \equiv \sum_{i=1}^8 r_i \lambda_i  \ \ , \ \ r_i \in Z^+ .
\eqno(II.6) $$
$Z^+$ here is the set of positive integers including zero. It is clear that
this last relation turns out to be
$$ \Lambda^+ \equiv \sum_{i=1}^8 q_i \sigma_i  \ \ , \ \ q_i \in Z^+ .
\eqno(II.7) $$
in view of (II.5) and hence $E_8$ $\leftrightarrow$  $A_8$. By comparison
between (II.6) and (II.7), note here that elements of $\Sigma(\lambda_1)$
are dominant weights for $A_8$ but not for $E_8$.

It is clear that we only need here to know the weights of the sets
$\Sigma(\lambda_i)$ for i=1,2, .. 8 \ explicitly. For instance,
$$ \Sigma(\lambda_1) = ( \sigma_1 + \sigma_8 \ , \ \sigma_3 \ , \ \sigma_6 ) $$
for which we have the decomposition
$$ \Pi(\lambda_1) = \Pi(\sigma_1+\sigma_8) \oplus \Pi(\sigma_3)
\oplus \Pi(\sigma_6) \eqno(II.8) $$
of 240 roots of $E_8$ Lie algebra. Due to permutational lemma given in ref(7),
$A_8$ Weyl orbits here are known to have the weight structures
$$ \eqalign{
\Pi(\sigma_1+\sigma_8) &= (\mu_{I_1}+\mu_{I_2}+\mu_{I_3}+\mu_{I_4}+
\mu_{I_5}+\mu_{I_6}+\mu_{I_7}+\mu_{I_8}) \cr
\Pi(\sigma_3) &= (\mu_{I_1}+\mu_{I_2}+\mu_{I_3})  \cr
\Pi(\sigma_6) &= (\mu_{I_1}+\mu_{I_2}+\mu_{I_3}+\mu_{I_4}+\mu_{I_5}+\mu_{I_6}) }
\eqno(II.9) $$
where all indices are permutated over the set (1,2, .. 9) providing no two of
them are equal. Note here by (II.4) that
$$ \eqalign{
\sigma_1+\sigma_8 &= \mu_1+\mu_2+\mu_3+\mu_4+\mu_5+\mu_6+\mu_7+\mu_8 \cr
\sigma_3 &= \mu_1+\mu_2+\mu_3  \cr
\sigma_6 &= \mu_1+\mu_2+\mu_3+\mu_4+\mu_5+\mu_6 }  \eqno(II.10)\ . $$
The formal similarity between (II.9) and (II.10) is a resume of the first
permutational lemma. Now, we are ready to state our second permutational
lemma:
\vskip 3mm
\par {\bf For a dominant weight $\lambda^+$, the set $\Sigma(\lambda^+)$ of
$A_8$ dominant weights is specified by
$$ \Sigma(\lambda^+) = \sum_{i=1}^8 r_i \ \Sigma(\lambda_i). \eqno(II.11) $$
\par together with the equality of square lengths.}
\vskip 3mm
\noindent In addition to $\Sigma(\lambda_1)$ given above, the other 7 sets
$\Sigma(\lambda_i)$ have respectively 7,15,27,35,17,5 and 11 elements for
i=2,3, .. 8 and they are given in appendix(1). It is therefore clear that
the weight decomposition of any $E_8$ Weyl orbit is now completely known
in terms of $A_8$ Weyl orbits in the presence of both of our lemmas.

\vskip 3mm
\noindent {\bf{III. \ \ DUALITY RULES FOR $A_8$}}
\vskip 3mm
In this section, we present some rules which we call {\bf $A_8$ Dualities}
in calculating $E_8$ cohomology. They are however similarly generalized for
Lie algebras other than $A_8$. It will be seen in the following that they
are of crucial importance in calculating $E_8$ cohomology relations higher
than degree 9.

We start by expressing an $A_8$ dominant weight $\sigma^+$ in the form
$$ \sigma^+ \equiv \sum_{i=1}^8 k_i \ \mu_i \ \ \ , \ \ \
k_1 \geq k_2 \geq ... \geq k_8 \geq 0  \ \   . \eqno(III.1) $$
To prevent repetitions, we reproduce here the main definitions and formulas
of ref(7) for $A_8$.
The eigenvalues of a Casimir operator of degree M then are known to be defined
by the aid of the formal definition
$$ ch_M(\sigma^+) \equiv \sum_{\mu \in \Pi(\sigma^+)} (\mu)^M \eqno(III.2) $$
for a Weyl orbit $\Pi(\sigma^+)$. Our way of calculation the right hand side
of (III.2) is given in appendix(2). To this end, we need to define the
following generators:
$$ \mu(M) \equiv \sum_{I=1}^9 (\mu_I)^M  \ \ . \eqno(III.3) $$
It is also convenient to define the following ones which we would like to call
{\bf K-generators}:
$$ K(M) \equiv \sum_{i=1}^8 (k_i)^M  \ \ . \eqno(III.4)  $$

We remark here by definition that $\mu(1) \equiv 0 $ and hence
$$ (\mu(1))^M \equiv 0 \ \ , \ \ M=1,2, .. ,9,10, ... . \eqno(III.5) $$
It can be readily seen that (III.5) is fulfilled for M=2,3, .. 9
without any other restriction. It gives rise however to the fact that ,
for $M \geq 10$, all the generators $\mu(M)$ are non-linearly depend on
the ones for M=2,3, .. 9. These non-linearities are clearly the reminiscents
of $A_8$ cohomology. We therefore call them {\bf $A_8$ Dualities}.
It will be  seen that the cohomology of $E_8$ Lie algebra will be provided
by these $A_8$ dualities.

The first example is
$$ \eqalign{
\mu(10) &\equiv {1  \over 8!} \ (25200 \ \mu(2)   \ \mu(8)
+ 19200 \ \mu(3)   \ \mu(7)
+ 16800 \ \mu(4)   \ \mu(6)                               \cr
&- 8400  \ \mu(2)^2 \ \mu(6)
- 13440 \ \mu(2)   \ \mu(3) \ \mu(5)
+ 8064  \ \mu(5)^2
+ 2100  \ \mu(2)^3 \ \mu(4)                               \cr
&- 5600  \ \mu(3)^2 \ \mu(4)
- 6300  \ \mu(2)   \ \mu(4)^2 + 2800  \ \mu(2)^2 \ \mu(3)^2
- 105   \ \mu(2)^5\ )  } \ \ . \eqno(III.6) $$
It is seen that $\mu(10)$ consists of p(10)=11 monomials coming from the
partitions of 10 into the set of numbers (2,3,4,5,6,7,8,9). We also
have p(8)=7, \ p(9)=8, \ p(11)=13, \ p(12)=19 and these are the maximum
numbers of monomials for corresponding degrees. We thus obtain the
following expressions:
$$ \eqalign{
\mu(11) &\equiv {1  \over 362880} \ ( \
- 3465  \ \mu(2)^4 \ \mu(3) + 12320 \ \mu(2) \ \mu(3)^3 +
41580 \ \mu(2)^2 \ \mu(3) \ \mu(4)      \cr
&- 41580 \ \mu(3) \ \mu(4)^2 + 16632 \ \mu(2)^3 \ \mu(5) -
44352 \ \mu(3)^2 \ \mu(5)               \cr
&-99792 \ \mu(2) \ \mu(4) \ \mu(5) - 110880 \ \mu(2) \ \mu(3) \ \mu(6) +
133056 \ \mu(5) \ \mu(6)                \cr
&- 71280 \ \mu(2)^2 \ \mu(7) + 142560 \ \mu(4) \ \mu(7) +
166320  \ \mu(3) \ \mu(8) + 221760 \ \mu(2) \ \mu(9) \ )      } $$

\noindent and

$$ \eqalign{
\mu(12) &\equiv {1 \over 725760} \ ( \
322560 \ \mu(3) \ \mu(9) + 136080 \ \mu(2)^2 \ \mu(8) +
272160 \ \mu(4) \ \mu(8) + 248832 \ \mu(5) \ \mu(7)     \cr
&-60480 \ \mu(2)^3 \ \mu(6) - 80640 \ \mu(3)^2 \ \mu(6) +
120960 \ \mu(6)^2 - 72576 \ \mu(2)^2 \ \mu(3) \ \mu(5)  \cr
&-145152 \ \mu(3) \ \mu(4) \ \mu(5) + 17010 \ \mu(2)^4 \ \mu(4) -
34020   \ \mu(2)^2  \ \mu(4)^2                          \cr
&- 22680 \ \mu(4)^3 + 20160 \ \mu(2)^3 \ \mu(3)^2 +
4480 \ \mu(3)^4 - 945 \ \mu(2)^6  \ )              } $$

\vskip 3mm
\noindent {\bf{IV.DECOMPOSITIONS OF INVARIANT POLINOMIALS IN THE $A_8$ BASIS}}
\vskip 3mm
Let us start with the decomposition
$$ ch_8(\Lambda^+) \equiv \sum_{\alpha=1}^7 Q_\alpha(\Lambda^+) \ T(\alpha)
\eqno(IV.1) $$
where 7 generators $T(\alpha)$ signify monomials
$$ \mu(8) \ , \ \mu(2) \mu(6) \ , \ \mu(3) \mu(5) \ , \ \mu(4)^2 \ , \
\mu(4) \mu(2)^2 \ , \ \mu(3)^2 \mu(2) \ , \ \mu(2)^4  $$
which are known to exist because p(8)=7. One must stress in (IV.1) that
coefficients $Q_\alpha(\Lambda^+)$ are assumed to be defined by comparison
of (IV.1) with (III.2). These 7 monomials play a prominent role in expressing
eigenvalues of an 8th order Casimir operator of $E_8$ Lie algebra because they
allow us to define the following {\bf invariant polinomials}:
$$ P_\alpha(\Lambda^+) \equiv
{Q_\alpha(\Lambda^+) \over Q_\alpha(\lambda_1)} \
{dimR(\lambda_1) \over dimR(\Lambda^+)} \
P_\alpha(\lambda_1)   \eqno(IV.2) $$
where $dimR(\Lambda^+)$ is the dimension of representation $R(\Lambda^+)$.
An important notice is the fact that we do not need the Weyl dimension formula
here. This will be provided by orbital decomposition (II.1) providing the
sets $\Sigma(\lambda^+)$ are known for each particular subdominant
$\lambda^+$ of $\Lambda^+$. Let us recall from ref(7) that dimensions of $A_8$
Weyl orbits are calculated by counting permutations. In definition (IV.2) of
invariant polinomials, the fundamental representation $R(\lambda_1)$ of $E_8$
is taken to be reference representation, i.e. all our expressions for Casimir
eigenvalues are to be given by normalizing with respect to fundamental
representation.

Explicit calculations for these 7 invariant polinomials $P_\alpha(\Lambda^+)$
show that we can find only 2 different polinomials the following one of which
comes from the monomial $\mu(2)^4$:
$$  P_1(8, \Lambda^+) \equiv 729 \ {\it \Theta}(8, \Lambda^+) -
71757069294212 \ . \eqno(IV.3)  $$
Only the following one is obtained for all other monomials:
$$ \eqalign{
P_2(8, \Lambda^+) &\equiv  68580 \ {\it \Theta}(8, \Lambda^+)  \cr
&-42672 \ {\it \Theta}(6, \Lambda^+) \ {\it \Theta}(2, \Lambda^+)     \cr
&-42672 \ {\it \Theta}(5, \Lambda^+) \ {\it \Theta}(3, \Lambda^+)     \cr
&-13335 \ {\it \Theta}(4, \Lambda^+)^2                                \cr
&+13335 \ {\it \Theta}(4, \Lambda^+) \ {\it \Theta}(2, \Lambda^+)^2   \cr
&+17780 \ {\it \Theta}(3, \Lambda^+)^2 \ {\it \Theta}(2, \Lambda^+)   \cr
&-939 \ {\it \Theta}(2, \Lambda^+)^4                                  \cr
&+385526887200     }  \eqno(IV.4)  $$

The functions $ {\it \Theta}(M, \Lambda^+)$ can be considered here as
{\bf $A_8$ basis functions} which are defined by
$$ {\it \Theta}(M, \Lambda^+) \equiv \sum_{I=1}^9 ( \vartheta_I(\Lambda^+))^M  \ \ \ ,
\ \ \ M = 1,2, ...  \eqno(IV.5)   $$
where
$$ \vartheta_I = \kappa(\Lambda^+ + \rho_w,\mu_I) \ \ . \eqno (IV.6) $$
$\rho_w$ here is the Weyl vector of $E_8$ Lie algebra. We notice that $A_8$
dualities are valid exactly in the same way for basis functions
${\it \Theta}(M, \Lambda^+)$ because ${\it \Theta}(1, \Lambda^+) \equiv 0 $.
This highly facilitates the work by allowing us to decompose all invariant
polinomials $ P_\alpha(\Lambda^+) $ in terms of ${\it \Theta}(M,\Lambda^+)$'s
but only for M=2,3,..9.

As in the similar way with $A_8$ basis functions defined above, the two
polinomials $P_1$ and $P_2$  can be considered as {\bf $E_8$ basis functions}
in the sense that for any 8th order Casimir operator of $E_8$ the
eigenvalues can always be expressed as linear superpositions of these
$E_8$ basis functions. What is really significant here is the allowance
of obtaining the decompositions (IV.3) and (IV.4) with coefficients which
are constant for all irreducible representations of $E_8$ Lie algebra.
In other words, beside constant coefficients, $E_8$ characteristic is
reflected by $A_8$ basis functions.

$E_8$ cohomology manifests itself here by the fact that we have 2 polinomials
$P_1$ and $P_2$ as {\bf $E_8$ Basis functions} in spite of the fact that
we have 7 polinomials from the beginning. As will be summarized in appendix(3),
the same considerations lead us for degree 12 to 19 different polinomials which
are known to exist from the beginning. It is however seen that the cohomology
of $E_8$ dictates only 8 invariant polinomials for degree 12.

Careful reader could now raise the question that is there a way for a direct
comparison of our results in presenting the $E_8$ basis functions
$$ \eqalign{
&P_\alpha(8,\Lambda^+) \ for  \ \alpha = 1,2 \cr
&P_\alpha(12,\Lambda^+) \ for \ \alpha = 1,2 .. 8  } \ \ . $$
A simple and might be possible way for such an investigation is due to weight
multiplicity formulas which can be obtained from these polinomials. The
method has been presented in another work {\bf [10]} for $A_N$ Lie algebras
and it can be applied here just as in the same manner. This shows the
correctness in our conclusion that any Casimir operator for $E_8$ can be
expressed as linear superpositions of $E_8$ basis functions which are given
in this work. An explicit comparison has been given in our previous works
but only for 4th and 5th order Casimir operators of $A_N$ Lie algebras and
beyond these this does not seem to be tractable in practice.

As the final remark, one can see that the method presented in this paper
are to be extended in the same manner to cases $E_7$ and $G_2$ in terms of
their sub-groups $A_7$ and $A_2$.

\vskip 3mm
\noindent {\bf{REFERENCES}}
\vskip 3mm
[1] M.B.Green and J.H.Schwarz, {\bf Phys.Lett. 149B} (1984) 117

[2] J.H.Schwarz : Anomaly-Free Supersymmetric Models in Six Dimensions,
hep-th/9512053

[3] D.J.Gross, J.A.Harvey, E.Martinec and R.Rohm : {\bf Phy.Rev.Lett. 54} (1985) 502

[4] J. Thierry-Mieg : {\bf Phys.Lett. 156B} (1985) 199

J. Thierry-Mieg : {\bf Phys.Lett. 171B} (1986) 163

[5] R. Hermann : Chapter 10, Lie Groups for Physicists, (1966) Benjamin

[6] A. Borel and C. Chevalley : {\bf Mem.Am.Math.Soc. 14} (1955) 1

Chih-Ta Yen: Sur Les Polynomes de Poincare des Groupes de Lie
Exceptionnels, {\bf Comptes Rendue Acad.Sci.} Paris (1949) 628-630

C. Chevalley : The Betti Numbers of the Exceptional Simple Lie Groups,
Proceedings of the International Congress of Mathematicians, 2 (1952) 21-24

A. Borel : Ann.Math. {\bf 57} (1953) 115-207

A.J. Coleman : {\bf Can.J.Math 10} (1958) 349-356

[7] H.R.Karadayi and M.Gungormez : Explicit Construction of Casimir
Operators and Eigenvalues:I , hep-th/9609060

H.R.Karadayi and M.Gungormez : Explicit Construction of Casimir
Operators and Eigenvalues:II , physics/9611002, to be appear in
{\bf Jour. of Math. Phys.}

[8] V.G.Kac, R.V.Moody and M.Wakimoto ; On $E_{10}$, preprint

[9] Humphreys J.E: Introduction to Lie Algebras and Representation
Theory , Springer-Verlag (1972) N.Y.

[10] H.R.Karadayi ; Non-Recursive Multiplicity Formulas for $A_N$ Lie algebras,
physics/9611008

\hfill\eject

\vskip 3mm
\noindent {\bf{APPENDIX.1}}
\vskip 3mm
The Weyl orbits of $E_8$ fundamental dominant weights $\lambda_i$ \ (i=1,2, .. 8)
are the unions of those of the following $A_8$ dominant weights:
$$ \eqalign{
\Sigma(\lambda_2) \equiv  (
&2 \ \sigma_1 + \sigma_7 \ , \
\sigma_1 + \sigma_3 + \sigma_8 \ , \
\sigma_1 + \sigma_6 + \sigma_8 \ , \cr
&\sigma_2 + \sigma_4 \ , \ \sigma_2 + 2 \sigma_8 \ , \
\sigma_3 + \sigma_6 \ , \ \sigma_5 + \sigma_7 )    \cr
&\ \ \ \ \ \ \ \ \ \ \ \ \ \ \ \ \ \ \
\ \ \ \ \ \ \ \ \ \ \ \ \ \ \ \ \ \ \cr
\Sigma(\lambda_3) \equiv  (
&\sigma_3 + 3 \sigma_8 \ , \ \sigma_2 + \sigma_3 + 2 \sigma_8 \ , \
\sigma_2 + \sigma_6 + 2 \sigma_8 \ , \cr
&3 \sigma_1 + \sigma_6 \ , \
\sigma_4 + 2 \sigma_7 \ , \ \sigma_1 + \sigma_3 + \sigma_6 + \sigma_8 \ , \
2 \sigma_1 + \sigma_3 + \sigma_7 \ , \cr
&\sigma_1 + \sigma_5 + \sigma_7 + \sigma_8 \ , \
2 \sigma_1 + \sigma_6 + \sigma_7 \ , \
\sigma_1 + \sigma_2 + \sigma_4 + \sigma_8 \ , \cr
&\sigma_1 + 2 \sigma_4 \ , \ 2 \sigma_2 + \sigma_5 \ ,
\ 2 \sigma_5 + \sigma_8 \ , \ \sigma_2 + \sigma_4 + \sigma_6 \ , \
\sigma_3 + \sigma_5 + \sigma_7 )   \cr
&\ \ \ \ \ \ \ \ \ \ \ \ \ \ \ \ \ \ \
\ \ \ \ \ \ \ \ \ \ \ \ \ \ \ \ \ \ \
\ \ \ \ \ \ \ \ \ \ \ \ \ \ \ \ \ \ \cr
\Sigma(\lambda_4) \equiv (
&\sigma_4 + 4 \sigma_8 \ , \ 2 \sigma_3 + 3 \sigma_8 \ ,
\sigma_3 + \sigma_6 + 3 \sigma_8 \ , \cr
&\sigma_1 + \sigma_4 + 2 \sigma_7 + \sigma_8 \ , \
\sigma_2 + \sigma_3 + \sigma_6 + 2 \sigma_8 \ , \
\sigma_2 + \sigma_5 + \sigma_7 + 2 \sigma_8 \ , \cr
&\sigma_3 + 3 \sigma_7 \ , \
2 \sigma_2 + \sigma_4 + 2 \sigma_8 \ , \
4 \sigma_1 + \sigma_5 \ , \cr
&2 \sigma_1 + \sigma_3 + \sigma_6 + \sigma_7 \ , \
2 \sigma_1 + 2 \sigma_4 + \sigma_8 \ , \
\sigma_1 + 2 \sigma_2 + \sigma_5 + \sigma_8 \ , \cr
&\sigma_1 + 2 \sigma_5 + 2 \sigma_8 \ , \
3 \sigma_2 + \sigma_6 \ , \
3 \sigma_1 + \sigma_3 + \sigma_6 \ , \cr
&\sigma_3 + \sigma_4 + 2 \sigma_7 \ , \
2 \sigma_1 + \sigma_5 + 2 \sigma_7 \ , \
\sigma_1 + \sigma_2 + \sigma_4 + \sigma_6 + \sigma_8 \ , \cr
&3 \sigma_1 + 2 \sigma_6 \ , \
2 \sigma_1 + \sigma_2 + \sigma_4 + \sigma_7 \ , \
\sigma_1 + \sigma_3 + \sigma_5 + \sigma_7 + \sigma_8 \ , \cr
&\sigma_2 + \sigma_4 + \sigma_5 + \sigma_7 \ , \
3 \sigma_4 \ , \ \sigma_3 + 2 \sigma_5 + \sigma_8 \ , \
\sigma_1 + 2 \sigma_4 + \sigma_6 \ , \
2 \sigma_2 + \sigma_5 + \sigma_6 \ , \ 3 \sigma_5 ) \cr
&\ \ \ \ \ \ \ \ \ \ \ \ \ \ \ \ \ \ \
\ \ \ \ \ \ \ \ \ \ \ \ \ \ \ \ \ \ \
\ \ \ \ \ \ \ \ \ \ \ \ \ \ \ \ \ \ \cr
\Sigma(\lambda_5) \equiv (
&\sigma_5 + 5 \sigma_8 \ , \
\sigma_4 + \sigma_6 + 4 \sigma_8 \ , \
\sigma_2 + \sigma_4 + 2 \sigma_7 + 2 \sigma_8 \ , \cr
&2 \sigma_3 + \sigma_6 + 3 \sigma_8 \ , \
\sigma_2 + 4 \sigma_7 \ , \
\sigma_3 + \sigma_5 + \sigma_7 + 3 \sigma_8 \ , \cr
&\sigma_1 + \sigma_3 + 3 \sigma_7 + \sigma_8 \ , \
5 \sigma_1 + \sigma_4 \ , \
2 \sigma_3 + 3 \sigma_7 \ , \cr
&2 \sigma_1 + \sigma_4 + 3 \sigma_7 \ , \
3 \sigma_2 + \sigma_5 + 2 \sigma_8 \ , \
\sigma_2 + 2 \sigma_5 + 3 \sigma_8 \ , \cr
&\sigma_1 + 3 \sigma_2 + \sigma_6 + \sigma_8 \ , \
\sigma_1 + \sigma_3 + \sigma_4 + 2 \sigma_7 + \sigma_8 \ , \
2 \sigma_2 + \sigma_4 + \sigma_6 + 2 \sigma_8 \ , \cr
&4 \sigma_2 + \sigma_7 \ , \
\sigma_2 + \sigma_3 + \sigma_5 + \sigma_7 + 2 \sigma_8 \ , \
3 \sigma_2 + 2 \sigma_6 \ , \cr
&3 \sigma_1 + \sigma_3 + 2 \sigma_6 \ , \
3 \sigma_1 + 2 \sigma_4 + \sigma_7 \ , \
2 \sigma_1 + 2 \sigma_2 + \sigma_5 + \sigma_7 \ , \cr
&\sigma_1 + \sigma_2 + \sigma_4 + \sigma_5 + \sigma_7 + \sigma_8 \ , \
4 \sigma_1 + \sigma_3 + \sigma_5 \ , \
2 \sigma_1 + \sigma_2 + \sigma_4 + \sigma_6 + \sigma_7 \ , \cr
&\sigma_1 + \sigma_3 + 2 \sigma_5 + 2 \sigma_8 \ , \
3 \sigma_1 + \sigma_2 + \sigma_4 + \sigma_6 \ , \
\sigma_2 + 2 \sigma_4 + 2 \sigma_7 \ , \cr
&2 \sigma_1 + \sigma_3 + \sigma_5 + 2 \sigma_7 \ , \
2 \sigma_1 + 2 \sigma_4 + \sigma_6 + \sigma_8 \ , \
\sigma_1 + 2 \sigma_2 + \sigma_5 + \sigma_6 + \sigma_8 \ , \cr
&3 \sigma_4 + \sigma_6 \ , \
\sigma_1 + 2 \sigma_4 + \sigma_5 + \sigma_7 \ , \
2 \sigma_2 + 2 \sigma_5 + \sigma_7 \ , \
\sigma_2 + \sigma_4 + 2 \sigma_5 + \sigma_8 \ , \
\sigma_3 + 3 \sigma_5 )  } $$



$$ \eqalign{
\Sigma(\lambda_6) \equiv (
&\sigma_6 + 3 \sigma_8 \ ,
\ \sigma_4 + \sigma_7 + 2 \sigma_8 \ , \
3 \sigma_7 \ , \cr
&\sigma_2 + 2 \sigma_7 + \sigma_8 \ ,
\ \sigma_1 + \sigma_3 + 2 \sigma_7 \ , \
2 \sigma_2 + \sigma_6 + \sigma_8 \ , \cr
&\sigma_1 + 2 \sigma_2 + \sigma_7 \ , \
\sigma_2 + \sigma_4 + \sigma_7 + \sigma_8 \ , \ 3 \sigma_2 \ , \
3 \sigma_1 + \sigma_3 \ , \ \sigma_3 + \sigma_5 + 2 \sigma_8 \ , \cr
&2 \sigma_1 + \sigma_2 + \sigma_5 \ , \
\sigma_1 + \sigma_4 + \sigma_5 + \sigma_8 \ , \
2 \sigma_1 + \sigma_4 + \sigma_6 \ , \ 2 \sigma_4 + \sigma_7 \ , \cr
&\sigma_1 + \sigma_2 + \sigma_5 + \sigma_7 \ , \
\sigma_2 + 2 \sigma_5 ) \cr
&\ \ \ \ \ \ \ \ \ \ \ \ \ \ \ \ \ \ \
\ \ \ \ \ \ \ \ \ \ \ \ \ \ \ \ \ \ \
\ \ \ \ \ \ \ \ \ \ \ \ \ \ \ \ \ \ \cr
\Sigma(\lambda_7) \equiv (
&\sigma_7 + \sigma_8 \ , \ \sigma_2 + \sigma_7 \ , \
\sigma_1 + \sigma_2 \ , \ \sigma_4 + \sigma_8 \ , \ \sigma_1 + \sigma_5 ) \cr
&\ \ \ \ \ \ \ \ \ \ \ \ \ \ \ \ \ \ \
\ \ \ \ \ \ \ \ \ \ \ \ \ \ \ \ \ \ \
\ \ \ \ \ \ \ \ \ \ \ \ \ \ \ \ \ \ \cr
\Sigma(\lambda_8) \equiv (
&3 \sigma_8 \ , \
\sigma_3 + \sigma_7 + \sigma_8 \ , \
3 \sigma_1 \ , \
2 \sigma_2 + \sigma_8 \ , \cr
&\sigma_5 + 2 \sigma_8 \ , \
\sigma_1 + 2 \sigma_7 \ , \
\sigma_1 + \sigma_2 + \sigma_6 \ , \cr
&\sigma_1 + \sigma_4 + \sigma_7 \ , \
2 \sigma_1 + \sigma_4 \ , \
\sigma_2 + \sigma_5 + \sigma_8 \ , \
\sigma_4 + \sigma_5 ) }  $$

As an example of (II.11), let us construct the set $\Sigma(\lambda_1+\lambda_7)$
from $\Sigma(\lambda_1)$ and $\Sigma(\lambda_7)$ in view of our second lemma.
The lemma states that elements $ \sigma \in \Sigma(\lambda_1+\lambda_7)$
are to be chosen from 15 elements of
$\Sigma(\lambda_1) \oplus \Sigma(\lambda_7)$ providing the conditions
$$ \kappa(\sigma,\sigma) = \kappa(\lambda_1+\lambda_7,\lambda_1+\lambda_7) $$
In result, one has only the following 13 elements:
$$ \eqalign{
\Sigma(\lambda_1+\lambda_7) \equiv (
&\sigma_1 + \sigma_7 + 2 \sigma_8 \ ,          \cr
&\sigma_1 + \sigma_2 + \sigma_7 + \sigma_8 \ , \cr
&2 \sigma_1 + \sigma_2 + \sigma_8 \ ,          \cr
&\sigma_1 + \sigma_4 + 2 \sigma_8 \ ,          \cr
&\sigma_6 + \sigma_7 + \sigma_8 \ ,            \cr
&2 \sigma_1 + \sigma_5 + \sigma_8 \ ,          \cr
&\sigma_2 + \sigma_3 + \sigma_7 \ ,            \cr
&\sigma_1 + \sigma_2 + \sigma_3 \ ,            \cr
&\sigma_3 + \sigma_4 + \sigma_8 \ ,            \cr
&\sigma_2 + \sigma_6 + \sigma_7 \ ,            \cr
&\sigma_4 + \sigma_6 + \sigma_8 \ ,            \cr
&\sigma_1 + \sigma_3 + \sigma_5 \ ,            \cr
&\sigma_1 + \sigma_5 + \sigma_6 ) \ \ . } $$

\vskip 3mm
\noindent {\bf{APPENDIX.2}}
\vskip 3mm
Let us first borrow the following quantities from ref(7):
$$ \eqalign{
\Omega_8(\sigma^+) &\equiv \ 40320 \ K(8)  \ \mu(8) \ + \cr
&20160 \ \bigl( \  \cr
& \ 35 \ K(4,4)  \ \mu(4,4) +
14 \ K(5,3)  \ \mu(5,3) +
7  \ K(6,2)  \ \mu(6,2) +
2  \ K(7,1)  \ \mu(7,1) \ \bigr) \ + \cr
&40320 \ \bigl( \                    \cr
& \ 20 \ K(3,3,2)  \ \mu(3,3,2) +
15 \ K(4,2,2)  \ \mu(4,2,2) \ + \cr
& \ 5  \ K(4,3,1)  \ \mu(4,3,1) +
3  \ K(5,2,1)  \ \mu(5,2,1) +
2  \ K(6,1,1)  \ \mu(6,1,1) \ \bigr) \ + \cr
&13440 \ \bigl( \                        \cr
& \ 540 \ K(2,2,2,2)  \ \mu(2,2,2,2) +
30  \ K(3,2,2,1)  \ \mu(3,2,2,1) \  + \cr
& \ 40  \ K(3,3,1,1)  \ \mu(3,3,1,1) +
15  \ K(4,2,1,1)  \ \mu(4,2,1,1) +
18  \ K(5,1,1,1) \ \mu(5,1,1,1) \ \bigr) \ + \cr
&483840 \ \bigl( \                          \cr
& \ 3 \ K(2,2,2,1,1) \ \mu(2,2,2,1,1) + K(3,2,1,1,1) \ \mu(3,2,1,1,1) +
2 \ K(4,1,1,1,1)  \ \mu(4,1,1,1,1) \ \bigr) \ + \cr
&967680 \ \bigl( \                            \cr
& \ 3 \ K(2,2,1,1,1,1)  \ \mu(2,2,1,1,1,1) +
5 \ K(3,1,1,1,1,1)  \ \mu(3,1,1,1,1,1) \ \bigr) \ + \cr
&29030400 \ K(2,1,1,1,1,1,1)  \ \mu(2,1,1,1,1,1,1) \ + \cr
&1625702400 \ K(1,1,1,1,1,1,1,1)  \ \mu(1,1,1,1,1,1,1,1) } $$

$$ \eqalign{ \Omega_{12}(\Lambda^+) &\equiv 40320  \ K(12)  \ \mu(12) \ + \cr
&5040 \ \bigl( \                         \cr
&1848  \ K(6,6)  \ \mu(6,6) +
792   \ K(7,5)  \ \mu(7,5) +
495   \ K(8,4)  \ \mu(8,4) \ +  \cr
&220   \ K(9,3)  \ \mu(9,3) +
66    \ K(10,2)  \ \mu(10,2) +
12    \ K(11,1)  \ \mu(11,1) \ \bigr) \ + \cr
&95040 \ \bigl( \                       \cr
&1575  \ K(4,4,4)  \ \mu(4,4,4) +
210   \ K(5,4,3)  \ \mu(5,4,3) +
252   \ K(5,5,2)  \ \mu(5,5,2) \ + \cr
&280   \ K(6,3,3)  \ \mu(6,3,3) +
105   \ K(6,4,2)  \ \mu(6,4,2) +
42    \ K(6,5,1)  \ \mu(6,5,1) \ + \cr
&60    \ K(7,3,2)  \ \mu(7,3,2) +
30    \ K(7,4,1)  \ \mu(7,4,1) +
45    \ K(8,2,2)  \ \mu(8,2,2) \ + \cr
&15    \ K(8,3,1)  \ \mu(8,3,1) +
5     \ K(9,2,1)  \ \mu(9,2,1) +
2     \ K(10,1,1)  \ \mu(10,1,1) \ \bigr) \ + \cr
&95040 \ \bigl( \                         \cr
&11200  \ K(3,3,3,3)  \ \mu(3,3,3,3) \ + \cr
&700    \ K(4,3,3,2)  \ \mu(4,3,3,2) +
1050   \ K(4,4,2,2)  \ \mu(4,4,2,2) \ + \cr
&350    \ K(4,4,3,1)  \ \mu(4,4,3,1) +
420    \ K(5,3,2,2)  \ \mu(5,3,2,2) \ + \cr
&280    \ K(5,3,3,1)  \ \mu(5,3,3,1) +
105    \ K(5,4,2,1)  \ \mu(5,4,2,1) \ + \cr
&168    \ K(5,5,1,1)  \ \mu(5,5,1,1) +
630    \ K(6,2,2,2)  \ \mu(6,2,2,2) \ + \cr
&70     \ K(6,3,2,1)  \ \mu(6,3,2,1) +
70     \ K(6,4,1,1)  \ \mu(6,4,1,1) \ + \cr
&60     \ K(7,2,2,1)  \ \mu(7,2,2,1) +
40     \ K(7,3,1,1)  \ \mu(7,3,1,1) \ + \cr
&15     \ K(8,2,1,1)  \ \mu(8,2,1,1) +
10     \ K(9,1,1,1)  \ \mu(9,1,1,1) \ \bigr) \ + \cr
&380160 \ \bigl( \                          \cr
&1260  \ K(3,3,2,2,2)  \ \mu(3,3,2,2,2) \ + \cr
&420   \ K(3,3,3,2,1)  \ \mu(3,3,3,2,1) +
1890  \ K(4,2,2,2,2)  \ \mu(4,2,2,2,2) \ + \cr
&105   \ K(4,3,2,2,1)  \ \mu(4,3,2,2,1) +
140   \ K(4,3,3,1,1)  \ \mu(4,3,3,1,1) \ + \cr
&105   \ K(4,4,2,1,1)  \ \mu(4,4,2,1,1) +
189   \ K(5,2,2,2,1)  \ \mu(5,2,2,2,1) \ + \cr
&42    \ K(5,3,2,1,1)  \ \mu(5,3,2,1,1) +
63    \ K(5,4,1,1,1)  \ \mu(5,4,1,1,1) \ + \cr
&42    \ K(6,2,2,1,1)  \ \mu(6,2,2,1,1) +
42    \ K(6,3,1,1,1)  \ \mu(6,3,1,1,1) \ + \cr
&18    \ K(7,2,1,1,1)  \ \mu(7,2,1,1,1) +
18    \ K(8,1,1,1,1)  \ \mu(8,1,1,1,1) \ \bigr) \ + \cr
&570240 \ \bigl( \                           \cr
&56700  \ K(2,2,2,2,2,2)  \ \mu(2,2,2,2,2,2) \ + \cr
&1260   \ K(3,2,2,2,2,1)  \ \mu(3,2,2,2,2,1) +
280    \ K(3,3,2,2,1,1)  \ \mu(3,3,2,2,1,1) \ + \cr
&840    \ K(3,3,3,1,1,1)  \ \mu(3,3,3,1,1,1) +
315    \ K(4,2,2,2,1,1)  \ \mu(4,2,2,2,1,1) \ + \cr
&105    \ K(4,3,2,1,1,1)  \ \mu(4,3,2,1,1,1) +
420    \ K(4,4,1,1,1,1)  \ \mu(4,4,1,1,1,1) \ + \cr
&126    \ K(5,2,2,1,1,1)  \ \mu(5,2,2,1,1,1) +
168    \ K(5,3,1,1,1,1)  \ \mu(5,3,1,1,1,1) \ +  \cr
&84     \ K(6,2,1,1,1,1)  \ \mu(6,2,1,1,1,1) +
120    \ K(7,1,1,1,1,1)  \ \mu(7,1,1,1,1,1) \ \bigr) \ + \cr
&79833600 \ \bigl( \                                  \cr
&90  \ K(2,2,2,2,2,1,1)  \ \mu(2,2,2,2,2,1,1) \ + \cr
&9   \ K(3,2,2,2,1,1,1)  \ \mu(3,2,2,2,1,1,1) +
8   \ K(3,3,2,1,1,1,1)  \ \mu(3,3,2,1,1,1,1) \ + \cr
&6   \ K(4,2,2,1,1,1,1)  \ \mu(4,2,2,1,1,1,1) +
10  \ K(4,3,1,1,1,1,1)  \ \mu(4,3,1,1,1,1,1) \ + \cr
&6   \ K(5,2,1,1,1,1,1)  \ \mu(5,2,1,1,1,1,1) +
12  \ K(6,1,1,1,1,1,1)  \ \mu(6,1,1,1,1,1,1) \ \bigr) \ + \cr
&479001600 \ \bigl( \                                  \cr
&36 \ K(2,2,2,2,1,1,1,1)  \ \mu(2,2,2,2,1,1,1,1) \ + \cr
&10  \ K(3,2,2,1,1,1,1,1)  \ \mu(3,2,2,1,1,1,1,1) +
40  \ K(3,3,1,1,1,1,1,1)  \ \mu(3,3,1,1,1,1,1,1) \ + \cr
&15  \ K(4,2,1,1,1,1,1,1)  \ \mu(4,2,1,1,1,1,1,1) +
42  \ K(5,1,1,1,1,1,1,1)  \ \mu(5,1,1,1,1,1,1,1) \ \bigr) } $$

In all these expressions, the so-called K-generators are to be reduced
to the ones defined by (III.3) for which the parameters $k_i$ are determined
via (III.1) for a dominant weight $\sigma^+$ which we prefer to suppress
from K-generators. The reduction rules can be deduced from definitions given
also in ref(7). The left-hand side of (III.2) can thus be calculated from
$$ ch_M(\sigma^+) \equiv {1 \over 9!} \ dim\Pi(\sigma^+) \ \Omega_M(\sigma^+) $$
with which we obtain $E_8$ Weyl orbit characters. The dimension of a Weyl
orbit $\Pi(\sigma^+)$ is the number of its elements and we show this number by
$dim\Pi(\sigma^+)$. Once again, we stress that both explicit forms and also the
number of these weights are known due to permutational lemma given in ref(7).

\vskip 3mm
\noindent {\bf{APPENDIX.3}}
\vskip 3mm
We now give the results of our 12th degree calculations. Explicit dependences
on $\Lambda^+$ will be suppressed here. It will be useful to introduce the
following auxiliary functions in terms of which the formal definitions
of $E_8$ basis functions will be highly simplified:
$$ \eqalign{
{\cal W}_1(8) &\equiv 68580  \ {\it \Theta}(8)
- 42672  \ {\it \Theta}(2)  \ {\it \Theta}(6) \ -  \cr
&42672  \ {\it \Theta}(3)  \ {\it \Theta}(5)
- 13335  \ {\it \Theta}(4)^2    \ +      \cr
&13335  \ {\it \Theta}(2)^2  \ {\it \Theta}(4)
+ 17780  \ {\it \Theta}(2)  \ {\it \Theta}(3)^2
- 939    \ {\it \Theta}(2)^4   } $$
$$ \eqalign{
{\cal W}_2(8) &\equiv  76765890960  \ {\it \Theta}(8)
- 47741514624  \ {\it \Theta}(2)  \ {\it \Theta}(6)   \ -  \cr
&47569228416  \ {\it \Theta}(3)  \ {\it \Theta}(5)
- 14950629660  \ {\it \Theta}(4)^2   \ +    \cr
&14921466630  \ {\it \Theta}(2)^2  \ {\it \Theta}(4)
+ 19832476160  \ {\it \Theta}(2)  \ {\it \Theta}(3)^2
- 1050561847   \ {\it \Theta}(2)^4  }$$
$$ \eqalign{
{\cal W}_1(12) &\equiv 302400 \ {\it \Theta}(3) \ {\it \Theta}(9) -
56700   \ {\it \Theta}(4)  \ {\it \Theta}(8)     \ -    \cr
&51840   \ {\it \Theta}(5)  \ {\it \Theta}(7)
- 158400  \ {\it \Theta}(2)  \ {\it \Theta}(3)  \ {\it \Theta}(7)
+ 30240   \ {\it \Theta}(6)^2       \ -     \cr
&168000  \ {\it \Theta}(3)^2  \ {\it \Theta}(6)
+ 33264   \ {\it \Theta}(2)  \ {\it \Theta}(5)^2
- 80640   \ {\it \Theta}(3)  \ {\it \Theta}(4)  \ {\it \Theta}(5)  \ +  \cr
&16275   \ {\it \Theta}(4)^3
+ 92400   \ {\it \Theta}(2)  \ {\it \Theta}(3)^2  \ {\it \Theta}(4)
+ 19600   \ {\it \Theta}(3)^4   }$$
$$ \eqalign{
{\cal W}_2(12) &\equiv 42338419200   \ {\it \Theta}(3)  \ {\it \Theta}(9)
- 7938453600    \ {\it \Theta}(4)  \ {\it \Theta}(8)  \ - \cr
&250343238600  \ {\it \Theta}(2)^2  \ {\it \Theta}(8)
- 7258014720    \ {\it \Theta}(5)  \ {\it \Theta}(7) \ -  \cr
&22177267200   \ {\it \Theta}(2)  \ {\it \Theta}(3)  \ {\it \Theta}(7)
+ 4233841920    \ {\it \Theta}(6)^2     \ -         \cr
&23521344000   \ {\it \Theta}(3)^2  \ {\it \Theta}(6)
+ 156357159840  \ {\it \Theta}(2)^3  \ {\it \Theta}(6)   \ +          \cr
&4657226112    \ {\it \Theta}(2)  \ {\it \Theta}(5)^2
- 11290245120   \ {\it \Theta}(3)  \ {\it \Theta}(4)  \ {\it \Theta}(5)   \ +    \cr
&160591001760  \ {\it \Theta}(2)^2  \ {\it \Theta}(3)  \ {\it \Theta}(5)
+ 2278630200    \ {\it \Theta}(4)^3      \ +     \cr
&48089818350   \ {\it \Theta}(2)^2  \ {\it \Theta}(4)^2
+ 12936739200   \ {\it \Theta}(2)  \ {\it \Theta}(3)^2  \ {\it \Theta}(4)  \ -   \cr
&48806484300   \ {\it \Theta}(2)^4  \ {\it \Theta}(4)
+ 2744156800    \ {\it \Theta}(3)^4        \ -    \cr
&66618900600   \ {\it \Theta}(2)^3  \ {\it \Theta}(3)^2
+ 3440480295    \ {\it \Theta}(2)^6    } $$
$$ \eqalign{
{\cal W}_3(12) &\equiv 1976486400  \ {\it \Theta}(3)  \ {\it \Theta}(9)
- 370591200   \ {\it \Theta}(4)  \ {\it \Theta}(8)      \ +          \cr
&63622800    \ {\it \Theta}(2)^2  \ {\it \Theta}(8)
- 338826240   \ {\it \Theta}(5)  \ {\it \Theta}(7)      \ -           \cr
&1035302400  \ {\it \Theta}(2)  \ {\it \Theta}(3)  \ {\it \Theta}(7)
+ 197648640   \ {\it \Theta}(6)^2                 \ -         \cr
&1098048000  \ {\it \Theta}(3)^2  \ {\it \Theta}(6)
- 12136320    \ {\it \Theta}(2)^3  \ {\it \Theta}(6)     \ +         \cr
&217413504   \ {\it \Theta}(2)  \ {\it \Theta}(5)^2
- 527063040   \ {\it \Theta}(3)  \ {\it \Theta}(4)  \ {\it \Theta}(5)    \ +    \cr
&185512320   \ {\it \Theta}(2)^2  \ {\it \Theta}(3)  \ {\it \Theta}(5)
+ 106373400   \ {\it \Theta}(4)^3            \ -         \cr
&39822300    \ {\it \Theta}(2)^2  \ {\it \Theta}(4)^2
+ 603926400   \ {\it \Theta}(2)  \ {\it \Theta}(3)^2  \ {\it \Theta}(4)   \ +  \cr
&6366150     \ {\it \Theta}(2)^4  \ {\it \Theta}(4)
+ 128105600   \ {\it \Theta}(3)^4              \ -     \cr
&63571200    \ {\it \Theta}(2)^3  \ {\it \Theta}(3)^2
- 274935      \ {\it \Theta}(2)^6       }  $$
$$ \eqalign{
{\cal W}_4(12) &\equiv - 1501985020838400  \ {\it \Theta}(3)  \ {\it \Theta}(9)
+ 192772901311200   \ {\it \Theta}(4)  \ {\it \Theta}(8)     \ +      \cr
&2407922770302000  \ {\it \Theta}(2)^2  \ {\it \Theta}(8)
- 13295642434560    \ {\it \Theta}(5)  \ {\it \Theta}(7)     \ +   \cr
&760428342950400   \ {\it \Theta}(2)  \ {\it \Theta}(3)  \ {\it \Theta}(7)
- 156516673824000   \ {\it \Theta}(6)^2       \ +      \cr
&33565287369600    \ {\it \Theta}(2)  \ {\it \Theta}(4)  \ {\it \Theta}(6)
+ 883577458444800   \ {\it \Theta}(3)^2  \ {\it \Theta}(6)   \ -      \cr
&1515778400455200  \ {\it \Theta}(2)^3  \ {\it \Theta}(6)
- 53070803904384    \ {\it \Theta}(2)  \ {\it \Theta}(5)^2   \ +      \cr
&579544204861440   \ {\it \Theta}(3)  \ {\it \Theta}(4)  \ {\it \Theta}(5)
- 1696086939738240  \ {\it \Theta}(2)^2  \ {\it \Theta}(3)  \ {\it \Theta}(5)  \ -  \cr
&47654628701400    \ {\it \Theta}(4)^3
- 461057612469300   \ {\it \Theta}(2)^2  \ {\it \Theta}(4)^2  \ -   \cr
&463327486742400   \ {\it \Theta}(2)  \ {\it \Theta}(3)^2  \ {\it \Theta}(4)
+ 472701971331450   \ {\it \Theta}(2)^4  \ {\it \Theta}(4)    \ -     \cr
&111245008649600   \ {\it \Theta}(3)^4
+ 684206487048000   \ {\it \Theta}(2)^3  \ {\it \Theta}(3)^2
- 33351005297925    \ {\it \Theta}(2)^6      }   $$

It is first seen that the expression (IV.4) can be cast in the form
$$ P_2(8) \equiv {\cal W}_1(8) \ + \ 385526887200 \ \  . $$
Let us further define
$$ \eqalign{
\Delta_{12} &\equiv ( \ {\it \Theta}(2) - 620) \ ( \ - 105 \ {\it \Theta}(2)^5
+ 341250             \ {\it \Theta}(2)^4
- 443786280          \ {\it \Theta}(2)^3   \ + \cr
&288672359200       \ {\it \Theta}(2)^2
- 93922348435072     \ {\it \Theta}(2)
+ 12228055880335360  \  )  } $$
with the remark that the square length of $E_8$ Weyl vector is 620. At last,
8  basis functions of $E_8$ will be expressed as in the following:
$$ \eqalign{
P_1(12) &\equiv {\cal W}_1(12) \ + \ \cr
& {_{105} \over _{1392517035128}} \ ( \   \cr
&_{2327783} \ {\cal W}_2(8) \ {\it \Theta}(2)^2 \ + \ \cr
&_{1641651348800} \ {\cal W}_1(8) \ {\it \Theta}(2)^2 \ + \ \cr
&_{1853819288565353101504512} \ {\it \Theta}(2)^2 \ - \ \cr
&_{5646385058438400} \ {\cal W}_1(8) \ {\it \Theta}(2) \ - \ \cr
&_{2457714965901036308812800000} \ {\it \Theta}(2) \ + \ \cr
&_{1878213525838949376} \ {\cal W}_1(8) \ +  \ \cr
&_{474462162108792} \ P_{12}(0) \ +  \ \cr
&_{814849980464400425555898009600} \ ) \ \cr
& \ \ \ \ \ \ \ \ \ \ \ \ \          \cr
P_2(12) &\equiv {\cal W}_1(12) \ + \ \cr
& {_{105} \over _{6580376}} \ ( \ \cr
&_{11} \ {\cal W}_2(8) \ {\it \Theta}(2)^2 \ - \cr
&_{13946970} \ {\cal W}_1(8) \ {\it \Theta}(2)^2 \ - \ \cr
&_{717386789108493504} \ {\it \Theta}(2)^2 \ +  \ \cr
&_{2185025300} \ {\cal W}_1(8) \ {\it \Theta}(2) \ +  \ \cr
&_{951080970408987600000} \ {\it \Theta}(2) \ - \ \cr
&_{726826815792} \ {\cal W}_1(8) \ - \ \cr
&_{315043889595739569446400} \  )  \cr
& \ \ \ \ \ \ \ \ \ \ \ \ \         }  $$
$$ \eqalign{
P_3(12) &\equiv - {_{10742925608415} \over _{467309767}} \ \Delta_{12}
+ {_{6983349} \over _{10867669}}  \  P_1(12)
+ {_{3884320} \over _{10867669}}  \  P_2(12) \cr
& \ \ \ \ \ \ \ \ \ \ \ \ \ \                   \cr
P_4(12) &\equiv - {_{2898884687985} \over _{487195289}} \ \Delta_{12}
+ {_{39572311} \over _{237932583}}  \ P_1(12)
+ {_{198360272} \over _{237932583}} \ P_2(12)   \cr
& \ \ \ \ \ \ \ \ \ \ \ \ \ \                   \cr
P_5(12) &\equiv - {_{511567886115} \over _{1063875427}} \ \Delta_{12}
+ {_{2327783} \over _{173189023}}   \ P_1(12)
+ {_{170861240} \over _{173189023}} \ P_2(12)   \cr
& \ \ \ \ \ \ \ \ \ \ \ \ \ \                   \cr
P_6(12) &\equiv - {_{2557839430575} \over _{69599327}} \  \Delta_{12}
+ {_{11638915} \over _{11330123}}  \  P_1(12)
- {_{308792} \over _{11330123}}   \  P_2(12)   \cr
& \ \ \ \ \ \ \ \ \ \ \ \ \ \                   \cr
P_7(12) &\equiv - {_{17904876014025} \over _{1362158257}} \ \Delta_{12}
+ {_{11638915} \over _{31678099}}         \ P_1(12)
+ {_{20039184} \over _{31678099}}         \ P_2(12) \cr
& \ \ \ \ \ \ \ \ \ \ \ \ \ \                   \cr
P_8(12) &\equiv - {_{26924625585} \over _{9942761}} \ \Delta_{12}
+ {_{2327783} \over _{30753191}}    \ P_1(12)
+ {_{28425408} \over _{30753191}}   \ P_2(12)  }    $$

\vskip 3mm
\vskip 3mm

It is seen that the generator $\Delta_{12}$  plays the role of
{\bf a kind of cohomology operators} in the sense that 6 generators
$P_\alpha(12)$ \ (for $\alpha$ = 3,4, ... 8) will depend linearly on the
first 2 generators $P_1(12)$ and $P_2(12)$ modulo $\Delta_{12}$. It is
therefore easy to conclude that all our 8 generators
$P_\alpha(12)$ \ (for $\alpha$ = 1,2, .. 8)
are linearly independent due to the fact that there are no a linear
relationship among the generators $P_1(12)$ and $P_2(12)$
modulo $\Delta_{12}$.

\end